\documentclass[sigconf, screen, nonacm]{acmart}

\usepackage{soul,color}

\newcommand{\ourmodel}{OMAR-RQ}

\usepackage{xcolor}
\usepackage{booktabs}
\usepackage{arydshln} 
\usepackage{xcolor}   

\setcopyright{acmlicensed}
\copyrightyear{2025}
\acmYear{2025}
\acmDOI{XXXXXXX.XXXXXXX}
\acmConference[MM '25]{Proceedings of the 33th ACM International Conference on Multimedia}{October 27--31, 2025}{Dublin, Ireland}

\begin{document}


\title{OMAR-RQ: Open Music Audio Representation Model \\ Trained with Multi-Feature Masked Token Prediction}

\author{Pablo Alonso-Jim\'enez, Pedro Ramoneda, R. Oguz Araz, Andrea Poltronieri, Dmitry Bogdanov}
\affiliation{%
  \institution{Music Technology Group, Universitat Pompeu Fabra}
  \city{Barcelona}
  \country{Spain}
}
\email{{pablo.alonso, pedro.ramoneda, recepoguz.araz, andrea.poltronieri, dmitry.bogdanov}@upf.edu}

\renewcommand{\shortauthors}{Alonso-Jim\'enez et al.}

\begin{abstract}

Developing open-source foundation models is essential for advancing research in music audio understanding and ensuring 
access to powerful, multipurpose representations for music information retrieval. We present \ourmodel, a 
model trained with self-supervision via masked token classification methodologies using a large-scale dataset with over 330,000 hours of music audio.
We experiment with different input features and quantization options, and achieve state-of-the-art performance in music tagging, pitch estimation, chord recognition, beat tracking, segmentation, and difficulty estimation among open self-supervised models.
We open-source our training and evaluation pipelines and model weights, available at \url{https://github.com/mtg/omar-rq}.
\end{abstract}

\ccsdesc[500]{Information systems~Music retrieval}

\keywords{self-supervised learning, music representation learning}



\maketitle

\section{Introduction}

Currently, there is strong interest in building representation models that map raw audio to compact vectors capturing musical meaning and structure~\cite{mccallum2022supervised, alonso2023efficient, li2023mert, won2024foundation}.
These representations can be applied to a variety of music analysis tasks such as semantic tagging, music analysis, and retrieval. 
Researchers commonly fine-tune these models or utilize their embeddings to solve downstream tasks.
This approach allows for taking advantage of large-scale pre-training to save on data and compute requirements while gaining generalizability.
We argue that access to a strong, open representation is essential for researchers and practitioners to overcome technical barriers and develop efficient music analysis solutions and applications.

\begin{figure}
    \centering
    \includegraphics[width=1\linewidth]{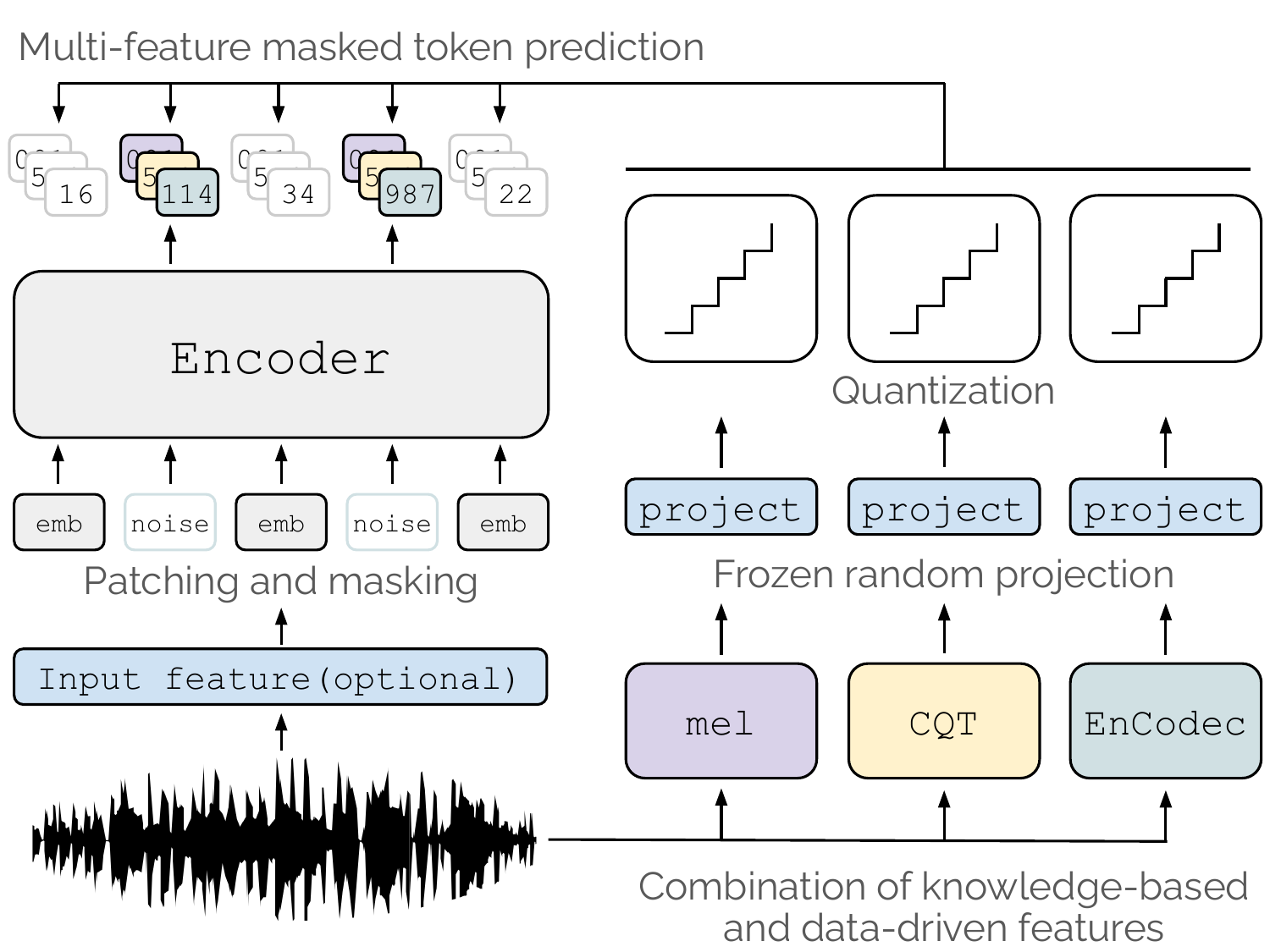}
    \caption{
    Diagram of the \ourmodel\ architecture.
    }
    \label{fig:enter-label}
\end{figure}

Current models learn representations directly from large-scale data using self-supervised learning (SSL) methods.
Popular techniques include contrastive learning~\cite{spijkervet2021contrastive} or masked prediction frameworks targeting continuous~\cite{huang2022masked} or discrete objectives derived from clustering~\cite{li2023mert} or random projections~\cite{won2024foundation}.
Although recent 
systems such as~\cite{won2024foundation} report strong performance in downstream tasks using these techniques, they lack openly distributed weights, which 
hinders reproducibility and slows the progress towards open music analysis solutions.

In this work, we present \ourmodel\, an SSL model trained with masked token classification building upon BestRQ, exploring various input and target features and quantization settings. We evaluate tokenization strategies applied to waveform, mel spectrograms, constant-Q transforms (CQT), neural encodings~\cite{defossez2022high}, and their combinations.
Additionally, we experiment with random codebooks and finite scale quantizer~\cite{mentzer2023finite} settings to optimize performance. 
We train and benchmark variations of \ourmodel\ in various music analysis tasks. 
Our results show that the proposed model performs competitively with recent closed-source systems.
To promote open-source music audio representation models, we release all pretrained model weights 
and source code for training and evaluation tools as part of an open research framework under the CC BY-NC-SA 4.0\footnote{\url{https://creativecommons.org/licenses/by-nc-sa/4.0/}} and AGPL-3.0\footnote{\url{https://www.gnu.org/licenses/agpl-3.0.en.html}} licenses.

\section{Pre-training methodology}
\label{sec:pretraining}

Reconstruction of partially masked input signals is an established SSL paradigm 
also adopted in audio~\cite{li2023mert}.
Different quantization strategies have been proposed in the audio domain, including k-means clustering~\cite{li2023mert}, random codebook projection~\cite{chiu2022self}, or finite scalar quantization~\cite {mentzer2023finite}.
Continuing this research line, we explore various tokenization approaches described hereby.

\begin{table}
\centering

\small
\begin{tabular}{ccccc}
\toprule
\multicolumn{3}{c}{Random codebook} & Pre-training & Downstream task \\
Num. & Codewords & Usage (\%) & valid. acc. & Tagging (mAP) \\
\midrule
1  & 32768  &  75.05\% & 0.257 & 0.461  \\
4  & 8192   & 96.46\% & 0.281 & \textbf{0.468} \\
16 & 2048   &   99.82\% & 0.339 &  0.465  \\
\bottomrule
\end{tabular}
\caption{
Effect of the codebook settings on the pre-training validation accuracy and the downstream task of auto-tagging. 
}
\label{tab:codebook_settings}
\end{table}

\textbf{BEST-RQ}. This method operates by projecting a sequence of audio input features (e.g., mel spectrograms) into a lower-dimensional space using a dense layer with random non-trainable weights~\cite{chiu2022self}.
The target labels are assigned by nearest-neighbor search in a codebook of random vectors, drawn from the uniform distribution.
The audio encoder is trained to predict labels corresponding to masked regions from the rest of the input signal.

\textbf{Multi-codebook.}
Previous works relying on audio quantization reported suboptimal performance associated with codebook underutilization~\cite{lancucki2020robust}.
In our work, we propose to utilize multiple codebooks initialized with different seeds.
This transforms our pre-training task into a multi-label classification problem, where each codebook provides a target label.


\textbf{Multi-feature.}
Since the goal of the multi-codebook method is to enrich the coverage of the target space,
we hypothesize that using different audio input features may further increase the representation power of the model.
We propose computing multiple targets by quantizing different input features.
In our work, we experiment with mel-spectrograms, CQT, EnCodec features~\cite{defossez2022high}, and the raw waveform as input features.

\textbf{Finite Scalar Quantization (FSQ).} 
We apply FSQ as a drop-in replacement for the random codebook assignments of BEST-RQ.
This method scales the low-dimensional projections to a given integer range by
$\left\lfloor \frac{L}{2} \right\rfloor \cdot \tanh(x)$, and quantized it via the round operator~\cite{mentzer2023finite}.


\section{Evaluation}
\label{sec:evaluation}

To evaluate the representations learned by the proposed and existing open SOTA models, we consider several downstream tasks covering both track-level and time-based musical notions.

\subsection{Downstream protocol and tasks }

Following common practice in the literature, we use 2-layer MLPs with 512 hidden units as probes to evaluate the proposed representations.
For track-level tasks, the embeddings extracted from the different models are averaged over the time dimension to obtain one vector representation per track.
For time-based tasks, the MLP operates on timestamps independently, and optional post-processing is applied following the standard practice for each task.\footnote{Our repository provides the hyperparameters used for each task.}

\textbf{Music tagging} is a track-level multi-label classification task of assigning labels related to musical genre, mood, era, or instrumentation.
We use the MagnaTagATune dataset containing 25,887 tracks with tag annotations
~\cite{law2009evaluation}, 
using an established split with tracks annotated by the 50 most frequent tags~\cite{won2024foundation}, and report results in terms of macro-averaged mean Average Precision (mAP).

\textbf{Pitch estimation} consists of estimating the fundamental frequency of musical signals.
We use the NSynth dataset containing 305,979 4-second single-note audio clips ranging from C-2 to B-7 (128 pitch classes)~\cite{nsynth2017}.
The results are reported in terms of accuracy.

\textbf{Chord recognition} is the harmony analysis task of 
assigning chord labels to successive audio frames.
We use four standard chord annotation datasets, Isophonics
~\cite{mauch2009beatles}
, Billboard
~\cite{burgoyne2011expert}
, Uspop 2002
~\cite{berenzweig2004large},
and RWC-Pop
~\cite{goto2002rwc}.
The data is split into 60\% training, 20\% validation, and 20\% test sets, using 30-second audio segments extracted with 50\% overlap.
The model predicts frame-level probabilities over 25 classes: 12 major, 12 minor, and a `none' label.
Evaluation is based on major/minor weighted accuracy using \texttt{mir\_eval}.

\begin{table}
\centering
\small
\begin{tabular}{llll}
\toprule
\multicolumn{2}{c}{Features} & \multicolumn{2}{c}{Downstream  tasks} \\
Input & Target & Tagging (mAP) & Pitch (acc.) \\
\midrule
mel & mel & \textbf{0.469} & 0.824 \\
cqt & cqt & 0.419 & 0.852 \\
enc & enc & 0.442 & \textbf{0.878} \\
\bottomrule
\end{tabular}
\caption{
Downstream performance for \ourmodel\ trained with mel-spectrograms, CTQ, or EnCodec embeddings.
}
\label{tab:reps_comp}
\end{table}

\textbf{Beat tracking} 
estimates the temporal locations of beats in a musical signal.
We use the HarmonixSet dataset~\cite{nieto2019harmonix} with 912 full tracks for training and validation, and the GTZAN dataset~\cite{marchand2015gtzan} with 1,000 30-second excerpts for testing.
We train and validate in 30-second segments extracted from the full tracks with a 50\% overlap.
For testing, we directly report the results on the GTZAN excerpts.
We used the F-measure metric as in previous work~\cite{won2024foundation}.


\begin{table}[b]
\centering
\small
\begin{tabular}{llll}
\toprule
Input & Target & Tagging & Pitch \\
\midrule
enc & enc  & 0.411 & 0.878 \\
enc & enc/mel & 0.445 & 0.898 \\
enc & enc/cqt & 0.433 & 0.892 \\
enc & enc/cqt/audio & 0.433 & 0.889 \\
enc & enc/mel/audio & 0.444 & 0.895 \\
enc & enc/mel/cqt & \textbf{0.446} & \textbf{0.900} \\
enc & enc/mel/cqt/audio & \textbf{0.446} & 0.878 \\  
\bottomrule
\end{tabular}
\caption{
Systematic comparison of target features in the downstream tasks of auto-tagging and pitch estimation.
}
\label{tab:multi_feature}
\end{table}

\textbf{Structure segmentation} task consists of dividing a music recording into non-overlapping sections and assigning each one a label such as `verse' or `chorus'. Following~\cite{won2024foundation}, we use two classifiers: one for frame-level functional classes and one for boundaries, with 200 ms frames. We divide the Harmonix Set~\cite{nieto2019harmonix} into 60\% training, 20\% validation, and 20\% test.
Table~\ref{tab:benchmark} shows the frame-wise accuracy
Unlike \cite{won2024foundation}, we do not use data augmentation or CTC loss. We release the split for fair comparisons.

\setlength\dashlinedash{1pt}
\setlength\dashlinegap{2pt}
\setlength\arrayrulewidth{0.3pt}

\begin{table*}[t]
\centering
\small
\begin{tabular}{lllllll:cccccc}
\toprule
Model & Size & RF & Input & Target & Quantizer & Rate & Tagging & Difficulty & Pitch & Chord & Beat & Structure \\
      & M    & s  &       &        &           & Hz 
      & \textit{mAP} & \textit{MSE} & \textit{acc.} & \textit{acc.} & \textit{F1} & \textit{acc.} \\
\midrule
MAEST & 86 & 30 & mel & genre labels & N/A & 3.9 
& \textbf{.491} & - & - & - & - & - \\
MERT & 330 & 5 & audio & cqt, enc & k-means & 75 
& .441 & 1.89 & .922 & .609 & \textbf{.868} & .393 \\
MusicFM FMA & 330 & 30 & audio & mel & RQ & 25 
& .448 & 1.78 & .857 & .644 & .847 & .635 \\
\midrule
\textbf{base} & 580 & 30 & mel & mel & RQ & 15.625 
& .482 & \textbf{1.65} & .892 & .657 & .783 & \textbf{.647} \\
\textbf{multi-codebook} & 580 & 30 & mel & mel & 4 $\times$ RQ & 15.625 
& .488 & 1.66 & .897 & .675 & .775 & .639 \\
\textbf{multi-feature} & 580 & 30 & audio & enc/mel/cqt/audio & 4 $\times$ RQ & 18.75 
& .467 & 1.76 & .938 & .734 & .833 & .623 \\
\textbf{high rate} & 580 & 30 & audio & enc/mel/cqt/audio & 4 $\times$ RQ & 25 
& .463 & 1.79 & .932 & .728 & .848 & .628 \\
\textbf{high rate FSQ} & 580 & 30 & audio & enc/mel/cqt/audio & 4 $\times$ FSQ & 25 
& .463 & 1.71 & \textbf{.940} & \textbf{.749} & .855 & .628 \\
\bottomrule
\end{tabular}
\caption{
Results on downstream tasks.
Bold font indicates our proposed OMAR-RQ variants.
}
\label{tab:benchmark}
\end{table*}

\textbf{Piano difficulty estimation} 
is a track-level regression task in which each music piece is assigned a difficulty score. We use the PianoSyllabus Dataset~\cite{ramoneda2025audio}, which includes graded annotations based on piano curricula. The probe is trained in the canonical split using an ordinal regression loss and reports mean squared error (MSE), following~\cite{ramoneda2025audio}.

\section{Experimental setup}
In this section, we experiment with different versions of \ourmodel\ to assess their performance in the selected downstream tasks.

\subsection{Pre-training dataset}

We pre-train our models in 330,000 hours of music audio extracted from 6.6 million YouTube videos.
Following previous work~\cite{alonso2022music, alonso2023efficient, lanzendorfer2023disco},  we relied on the Discogs\footnote{\url{https://www.discogs.com/}} 
music database to create our dataset.
Discogs releases monthly editorial metadata dumps under a CC0-BY license, including YouTube URLs.
This process favors building a corpus of high-quality and diverse music audio.\footnote{The Discogs database has 15 music genres categorized into 700 styles. 
Our companion website provides a complete list of the YouTube URLs used for pre-training.}

\subsection{Implementation details}

All our models use the Conformer architecture~\cite{gulati2020conformer}
considering two configurations: a \textit{base} model with 12 layers, 512 embedding dimensions, and 90 million parameters, and a \textit{large} model with 24 layers, 1024 embedding dimensions, and 580 million parameters following the Conformer XL configuration from ~\cite{zhang2020pushing}.
The \textit{base} architecture is used in our development experiments, and the \textit{large} architecture is used in our final models reported in Section~\ref{tab:benchmark}.
Both architectures use ROPE positional encoding~\cite{su2024roformer}, flash attention~\cite{Dao2022FlashAttentionFA}, and automatic mixed-precision training.\footnote{We used PyTorch Lightning's \texttt{bf16-mixed} precision.}
Additionally, we use DeepNorm ($\alpha=2.632$ and $\beta=0.022$) to prevent training collapse observed in the \textit{large} architectures~\cite{wang2024deepnet}.
All the models are trained for 400,000 steps, using cosine annealing with 30,000 warm-up steps and a maximum learning rate of $1 \times 10^{-4}$.
We use dropout and weight decay values of $0.2$ and $1 \times 10^{-2}$ for regularization.
Our models are trained with a combined batch size of 256 30-second samples using 4 Nvidia 64GB H100 GPUs for the \textit{small} version and 8 for the \textit{large} one. 
Due to the GPU memory constraints, the high-rate variants featured a combined batch size of 160.
Training takes over 3 days for our \textit{small} models and 5 days for the \textit{large} ones.

Our BEST-RQ configuration follows previous works~\cite{chiu2022self, won2024foundation}.
Our models use 30-second segments from which 60\% is masked in chunks of 0.4 seconds.
We experimented with Gaussian noise and waveform shuffling as our masking strategies, obtaining better results with the former method.
We use a single codebook with 8196 codewords of 16 dimensions in our \textit{base} configuration, and four codebooks with the same parameters in the \textbf{multi-codebook}, \textbf{multi-feature}, and \textbf{high rate} versions.
For the \textbf{high rate FSQ} version, we report the performance with 5 channels and 6 levels.

\section{Results}
\label{sec:results}

In this section, we show the results of the main experiments conducted in the development of \ourmodel.

\textbf{Multi-codebook settings.}
Table~\ref{tab:codebook_settings} compares three \textit{small} models targeting the same number of codewords arranged in 1, 4, and 16 parallel codebooks.
We observe that a larger codebook allows for more detailed targets but results in codebook underutilization.
On the other hand, many parallel codebooks allow for a more efficient use of the codebook at the expense of less accurate quantization.

\textbf{Comparison of features.} 
We assess the performance of different features both as input and training targets.
Particularly, we train \textit{small} architectures using mel-spectrograms, CQT, and EnCodec embeddings before the quantization stage~\cite{defossez2022high} as input and target features.
Table~\ref{tab:reps_comp} compares the models in auto-tagging and pitch estimation. 
Interestingly, the EnCodec embeddings perform higher than the CQT, which explicitly contains pitch class information.
Upon further inspection, we observed that the CQT tokenizer frequently assigned the same codeword to samples from the same instrument and close pitches at the higher octaves.
This resulted in poorer high-frequency performance.
Thus, we hypothesized that combining both features could enhance prediction performance.

\textbf{Multi-feature targets.} 
Table~\ref{tab:multi_feature} compares multiple \textit{small} architectures trained with a fixed input and a systematic combination of features as target.
Our results show that using multiple features as targets improves performance for all the cases except for the waveform, which tends to decrease the performance in pitch estimation.

\textbf{Benchmark with SOTA models.} We present a benchmark comparing \ourmodel\ with relevant SOTA models.
We consider MusicFM~\cite{won2024foundation}, which also follows a BEST-RQ approach and for which the authors provide open weights for one of its versions.
Also, we considered MERT~\cite{li2023mert}, with audio tokenization based on k-means.
Finally, we consider MAEST, an audio transformer trained on the same dataset as \ourmodel\ and targeting music style tags from Discogs suitable for tagging tasks~\cite{alonso2023efficient}.
We consider different variants of \ourmodel\ accounting for the previous experiments.
The main differences, along with the performance result, are summarized in Table~\ref{tab:benchmark}.
\textbf{\ourmodel\ multi-codebook} obtains the best performance in music tagging among the open SSL models, although MAEST achieves higher performance thanks to its supervised training and task alignment with the Discogs genre tags.
For pitch estimation and chord recognition, \textbf{\ourmodel\ high-rate} achieves the highest performance among the compared models, suggesting that the additional temporal resolution and more musically-aware frequency representation benefit these tasks.
For beat tracking, we observe that the results are favoured by a higher temporal resolution, resulting in MERT
performing the best, followed by the high-rate variants of our models. 
For structure segmentation, we observe small variations among our models.
Interestingly, variants with mel-spectrogram targets perform better for this task.
In the current probe evaluation setup, the structure recognition task tends to perform better with the same representations as auto-tagging instead of beat-tracking (which is also a time-based task).
This suggests that representations focused on temporal details are not as useful as those that perform well on the classification of semantic categories for this task.
Lastly, the \textit{base} model achieved the highest MSE value in difficulty estimation, following the same trend as auto-tagging.

We observe two trends in the selected downstream tasks: High-level music information tasks, where models without tonal information and lower representation rates tend to perform better, and low-level tasks that seem to benefit from higher temporal resolution rates and tonal representations.
Finally, \ourmodel\ achieves the highest performance among the open models considered in our validation, although other works have reported higher results in
undisclosed settings.

\section{Conclusion}

In this work, we build upon the BEST-RQ paradigm to create a new generation of open-source SSL music audio representation models for music. These models can be applied to various music analysis tasks, and we envision their usage by music information retrieval researchers and practitioners and beyond. We expect them to foster further advancements in representation learning research and downstream music audio analysis tasks, with potential for developing final applications.
To develop our models, we experimented with SOTA SSL methodologies based on masked token prediction to build models with a higher overall performance across a range of benchmark music audio analysis tasks.
Our technical contributions include experimental results utilizing parallel codebooks, multi-feature targets, and other quantization strategies. 




\begin{acks}
This work is supported by “IA y Música: Cátedra en Inteligencia Artificial y Música” (TSI-100929-2023-1) funded by the Secretaría de Estado de Digitalización e Inteligencia Artificial and the European Union-Next Generation EU, under the program Cátedras ENIA. We thankfully acknowledge the computer resources at MareNostrum and the technical support provided by Barcelona Supercomputing Center (IM-2024-2-0034).
\end{acks}

\bibliographystyle{ACM-Reference-Format}
\bibliography{acm-bib}

\end{document}